\documentclass[a4paper]{jpconf}

\usepackage{graphicx}
\usepackage{amsfonts}
\usepackage{amssymb}
\usepackage{bm}
\usepackage{dcolumn}
\usepackage{epsfig}
\usepackage[T1]{fontenc}
\usepackage{graphicx}
\usepackage{graphics}
\usepackage{hyperref}
\usepackage[latin1]{inputenc}
\usepackage{latexsym}
\usepackage{multirow}

\newcommand{\rsu}{r^{\ast}}
\newcommand{\rsup}{r^{\ast}_{p}}

\newcommand\be{\begin{equation}}
\newcommand\ba{\begin{eqnarray}}
\newcommand\ee{\end{equation}}
\newcommand\ea{\end{eqnarray}}

\newcommand{\mb}[1]{\mbox{\boldmath $#1$}}
\newcommand{\salto}[1]{\left[\,#1\,\right]^{}_{p}}

\newcommand{\met}{\mbox{g}}

\newcommand{\singu}{{\mbox{\tiny S}}}
\newcommand{\regu}{{\mbox{\tiny R}}}

\newcommand{\LL}{{\mbox{\tiny L}}}
\newcommand{\RR}{{\mbox{\tiny R}}}

\begin{document}

\title{Time-domain modelling of Extreme-Mass-Ratio Inspirals for the Laser Interferometer Space Antenna}

\author{Priscilla Canizares$^{1}$, and Carlos F. ~Sopuerta$^{2}$}

\address{$^{1,2}$ Institut de Ci\`encies de l'Espai (CSIC-IEEC). 
Facultat de Ci\`encies, Campus UAB, Torre C5 parells. Bellaterra, 08193 Barcelona, Spain.}

\ead{$^{1}$ pcm@ieec.uab.es, $^{2}$ sopuerta@ieec.uab.es}

\begin{abstract}
When a stellar-mass compact object is captured by a supermassive black hole
located in a galactic centre, the system losses energy and angular momentum by
the emission of gravitational waves. Subsequently, the stellar compact object
evolves inspiraling until plunging onto the massive black hole. These EMRI
systems are expected to be one of the main sources of gravitational waves for
the future space-based Laser Interferometer Space Antenna (LISA). However, the
detection of EMRI signals will require of very accurate theoretical templates
taking into account the gravitational self-force, which is the responsible of
the stellar-compact object inspiral. Due to its potential applicability on
EMRIs, the obtention of an efficient method to compute the scalar self-force
acting on a point-like particle orbiting around a massive black hole is being
object of increasing interest. We present here a review of our
time-domain numerical technique to compute the
self-force acting on a point-like particle and we show its suitability to deal
with  both circular and eccentric orbits.
\end{abstract}

\section{Introduction}
It is known that the majority of the galactic nuclei harbour a (super)massive black
hole (MBH).  Such a MBH can gravitationally capture a stellar-mass
compact object (SCO), with mass in the range $m = 1-50 M^{}_{\odot}$, which eventually
will perform a slow inspiral due to the emission of gravitational radiation until it
plunges onto the MBH.   These events are usually known as Extreme-Mass-Ratio Inspirals.
EMRIs with MBHs with masses in the range $M= 10^4-10^7 M_{\odot}$ are on the main
targets of the future space-based gravitational-wave observatory LISA.   One main 
issue is that EMRI signals will be buried in the LISA instrumental noise and the
gravitational-wave foreground produced by compact galactic binaries inside the LISA band.
In order to pull out these signals from the noise, we will have to cross-correlate the 
LISA data stream with a bank of EMRI waveform templates. These templates should be accurate 
enough to capture the physical relevant parameters of the system.   For this reason, we
need to include very precisely the gravitational \emph{back-reaction} of the SCO on its
own trajectory when modelling EMRIs. Due to the extreme mass ratios involved, an EMRI can 
be modeled within the framework of black hole perturbation theory, where the SCO is pictured 
as a point particle moving in the spacetime geometry of the MBH and the gravitational back 
reaction is described as the action of a local force, the \emph{self-force}. 

As a testbed for the development of numerical techniques to compute the self-force, we use
a simplified EMRI system, in which the SCO is a scalar charged particle, with charge $q$, which orbits a 
non-rotating Schwarzschild MBH~\cite{Poisson:2004lr} and whose inspiral is driven by 
a self-force generated by a scalar field, $\Phi$.  This simplified model contains all the essential
ingredients of the gravitational problem.
The scalar field $\Phi$ generated by the particle satisfies a wave equation 
in the MBH geometry, and the {\em scalar} self-force acting on the particle is
given by: $F^{\mu} = q (\met^{\mu\nu} + u^{\mu}u^{\nu}) \left.
\left(\nabla^{}_{\nu}\Phi\right) \right|^{}_{\gamma}$, where $u^{\mu} = dz^{\mu}/d\tau$ is the particle's velocity
, and $\tau$ is proper time.

The spherical symmetry of the MBH spacetime implies that the retarded field can be decomposed into 
decoupled scalar spherical harmonics modes: 
\begin{equation}
\Phi(x)= \sum_{\ell = 0}^{\infty}\sum_{m = -\ell}^{\ell}\Phi^{\ell m}(t,r)Y^{\ell m}(\theta,\phi)\;.
\end{equation}
However, due to the singular character of the point particle, the sum over all
the retarded field modes diverges at the particle location and, in order to
obtain a self-force with a truly physical meaning, the field modes must be
regularised. In order to do so, we use the {\em mode sum} regularisation
scheme~\cite{Barack:1999wf,Barack:2000eh,Barack:2001bw,Barack:2002mha,
Detweiler:2002gi,Haas:2006ne}, which gives an analytical expression for the
singular contribution of the retarded field, $\Phi^{\singu}$, at the particle
location. Then, by subtracting the singular field modes from the
retarded ones we obtain a smooth and differentiable field, $\Phi^{\regu} =
\Phi - \Phi^{\singu}$, which generates the scalar self-force:$\textit{F}^{\;\mu} = q (g^{\mu\;
\alpha} + u^{\mu} u^{\alpha} )\nabla_{\alpha}\Phi^{\regu}$. Here, we  
review our work on the development of new techniques for the numerical computation
of the retarded field,
$\Phi$.  We also show the results we have obtained for the self-force after applying the 
mode-sum scheme.

\section{Avoiding singularities: The Particle-without-Particle Formulation}
\label{pwpformulation}

\begin{figure*}
\centering
\includegraphics[width=0.80\textwidth]{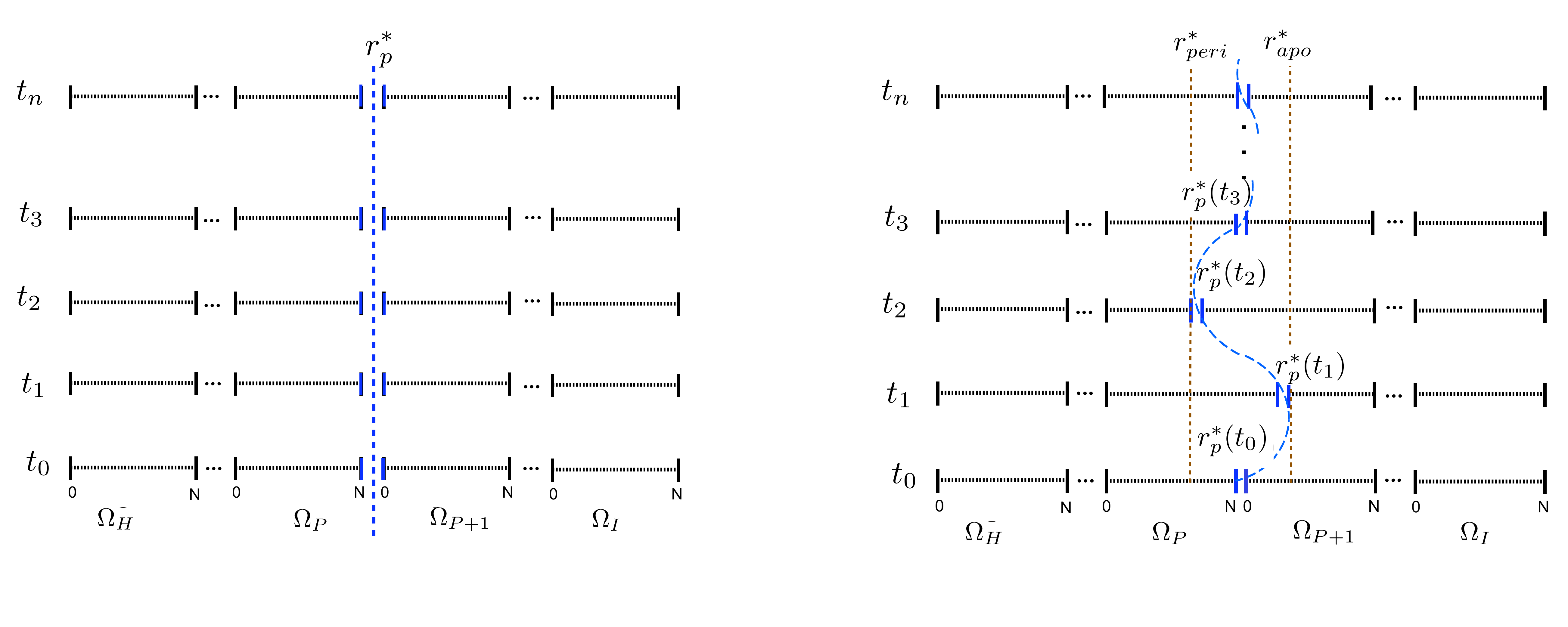}
\caption{One-dimensional spatial grid for circular orbits (left) and eccentric orbits (right).}
\label{multidomain}
\end{figure*}

In order to avoid the presence of the singularity associated with the point particle and 
all the related issues related with its numerical resolution we
have proposed a new time-domain computational technique: {\em the
Particle-without-Particle} (PwP) scheme.  This
technique was introduced in~\cite{Canizares:2008dp,Canizares:2009ay} for the case of a charged
scalar point particle orbiting in a circular orbit around a non-rotating MBH. In
a subsequent work~\cite{Canizares:2010yx}, the procedure was adapted to the
case of eccentric orbits (see also~\cite{Canizares:2011kw}).

\begin{figure}[htp]
\centering  
\includegraphics[width=1.0\textwidth]{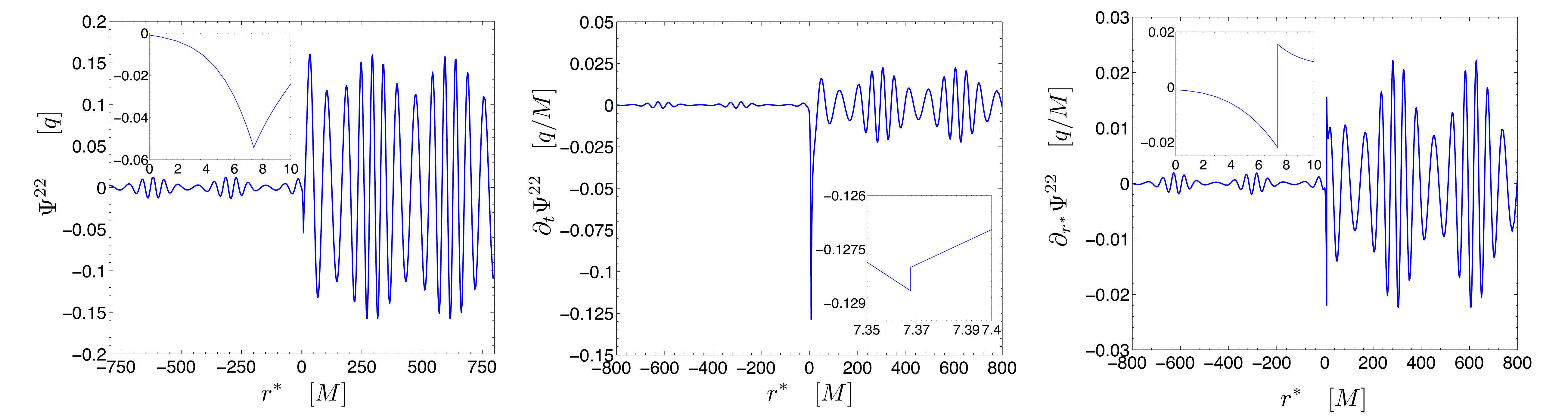}
\caption{Snapshots of the evolution of the scalar field variable
$\Psi = r\,\Phi$ (left), its time derivative $\partial_t \Psi$ (center), and its radial
derivative $\partial_{r*}\Psi$ (right), for a particle in an eccentric orbit with
orbital parameters $(e = 0.2, p = 7\, M)$.  The internal frames show the ability of
our method to resolve the different variables at the particle location, despite of 
the jumps in the time and radial derivatives.  The snapshots have been
taken after an evolution time $t^{}_{\rm\tiny final} - t^{}_{\rm\tiny initial} = 860 M$. }\label{scalar_fields}
\end{figure}

The main feature of our numerical scheme consists in splitting the one-dimensional 
spatial computational domain, parameterized in terms of the radial tortoise coordinate $\rsu$, 
into two regions or subdomains, one at the left and one at the right of the particle: 
$\rsu \in [-\infty, \infty] =[-\infty,\rsup]\cup [\rsup,\infty]$, where $\rsup$ denotes the 
particle radial location.   Based on this division of the 
computational domain, all the variables associated with our problem can be written as: 
$\mb{U}(t,\rsu) = \mb{U}^{}_{-}(t,\rsu)\Theta(\rsup(t)-\rsu)+\mb{U}^{}_{+}(t,\rsu)\Theta(\rsu - \rsup(t))\,,$
where $\Theta$ is the Heaviside step function. In this way, by defining the jump
of the variables 
as $\salto{\mb{U}} \equiv \mathop{\lim }\limits_{\rsu \to
\rsu_{p}}\mb{U}^{}_{+}(t,\rsu) - \mathop{\lim }\limits_{\rsu \to
\rsu_{p}}\mb{U}^{}_{-}(t, \rsu)$ 
(see~\cite{Sopuerta:2005gz,Canizares:2009ay,Canizares:2010yx})
we can obtain the jumps (matching conditions) at the particle location.
Using this information we evolve the $1+1$ wave-type equations for each mode inside each subdomain 
(which are homogeneous equations thanks to the PwP scheme).  The solutions
obtained are communicated through the boundaries by enforcing the analytic jumps.
We have developed two practical methods for enforcing the jumps:  
(i) The \emph{penalty} method, and (ii) the direct communication of the characteristic fields
(see~\cite{Canizares:2010yx}).   This is the essence of the PwP scheme, i.e. replacing
the particle by a set of boundary conditions. 

In our implementation of the PwP scheme, the spatial domain is discretized using
the PseudoSpectral Collocation (PSC) method.  In this way, each subdomain is discretized
independently using a {\em Lobatto-Chebyshev} grid and the variables are
expanded in a basis of Chebyshev polynomials (see~\cite{Boyd,Canizares:2009ay}).
One of the most interesting properties of the PSC method is that it provides exponential
convergence for smooth functions.  The implementation for the circular case 
has no special difficulties as the particle is always at the same radial position
(see Figure~\ref{multidomain}) and the domain structure is static and given by  
$\Omega = \bigcup^{D}_{a=1}\Omega^{}_a$, where  $\Omega^{}_a = \left[ \rsu_{a,\LL},
\rsu_{a,\RR}\right]$, and the particle is placed at the interface of two subdomains, say
$\Omega^{}_{a}$ and $\Omega^{}_{a+1}$,  with $\rsu_{p} =\rsu_{a,\RR} =
\rsu_{a+1,\LL} = \mbox{const.}$  However in the eccentric case, the radial position of the 
particle changes with time.  In this case, we keep the particle at the interface
by using a time-dependent mapping between the spectral domain, $[-1,1]$, and the corresponding
physical subdomains, so that the particle is always mapped to the boundaries of the
spectral domain (see Figure~\ref{multidomain}).

\begin{center}
\begin{table}[h]
\centering
\caption{Self-Force components obtained for $\ell_{max}= 20$ and $24$ subdomains} 
\begin{tabular}{@{}l*{15}{l}{l}}
\br
&Circular Orbit Case& (e = 0.0, p = 6 M)&\\
\mr
$\partial_t\Phi^R$ & ~~~$\partial_r\Phi^R$ & $\partial_{\phi}\Phi^R$\\
\mr
$3.608899\cdot10^{-4}$   & ~~~ $1.677117\cdot10^{-4}$ &  $-5.304114 \cdot10^{-3}$ \\
\mr
\mr
&Eccentric Orbit Case& (e = 0.2, p = 7 M)&\\
\mr
$\partial_t\Phi^R$ &~~~ $\partial_r\Phi^R$ & $\partial_{\phi}\Phi^R$\\
\mr
$4.278105\cdot10^{-4} $ &~~~ $1.946249\cdot10^{-4}$& $-5.961118\cdot10^{-3}$\\
\br			\label{self_force}
\end{tabular}
\end{table}
\end{center}

\section{Discussion}\label{Discussion}
Here we have summarized our work on the development of new time-domain scheme for the 
computation of the self-force~\cite{Canizares:2009ay,Canizares:2010yx}, the PwP scheme.
We have shown it is a suitable scheme by presenting results on the simplified case
of a scalar point particle in circular and eccentric orbits around a non-rotating
MBH.  In Figure~\ref{scalar_fields} we show snapshots of the evolution of the
different variables used in our implementation both for circular and eccentric 
orbits.  This figure shows that the computational scheme can resolve in a very
smooth way the retarded field and its derivatives at the particle location.
In particular it can resolve jumps in the derivatives of the scalar field, due to
the particle (this is how the particle appears in our PwP scheme). Then, by applying
the mode-sum regularization scheme we obtain the self-force components. In Table~\ref{self_force} 
we show the self-force results obtained for the particle in a circular orbit, $(e = 0.0, p = 6)$, 
and also in a eccentric orbit with orbital parameters: $(e = 0.2, p = 7 M)$.  
In the circular case,  the radial component of the self-force, the only one that needs 
to be regularized, agrees within a $1\cdot10^{-4}\%$ with the results obtained 
by~\cite{DiazRivera:2004ik} in the frequency domain.
In Table~\ref{self_force} we also present results obtained for the
eccentric orbit just mentioned.  These results have been obtained
by employing between 20 to 24 subdomains and 50 collocation points per domain.  
The average time for a full self-force calculation with $\ell_{max} = 20$ (which involves the 
calculation of 231 harmonic modes) in a computer with two Quad-Core
Intel Xeon processors at 2.27GHz is around 10 to 15 minutes.  See~\cite{Canizares:2011kw} for a detailed description
of how to tune the method for performance and efficiency.  In another recent work~\cite{Jaramillo:2011gu}
we discussed the question of initial conditions and how to avoid the presence of spurious
solutions that can affect self-force computations.


\section*{Acknowledgments}
PCM is supported by a predoctoral FPU fellowship of the Spanish Ministry of Science and Innovation.
CFS acknowledges support from the Ram\'on y Cajal Programme of the Ministry of Education and Science 
of Spain and by a Marie Curie
International Reintegration Grant (MIRG-CT-2007-205005/PHY) of the European Community (FP7).
We acknowledge the
computational resources provided by the Centre de Supercomputaci\'o de Catalunya
and the Centro de Supercomputaci\'on de Galicia (ICTS 121 and 175).

\section*{References}
\providecommand{\newblock}{}

\end{document}